\documentclass[
reprint,
showpacs,
nofootinbib,
 amsmath,amssymb,
 aps,
pra,
floatfix,
]{revtex4-1}

\usepackage{graphicx}
\usepackage{dcolumn}
\usepackage{bm}
\usepackage[colorlinks=true, citecolor=blue, urlcolor=blue ]{hyperref}
\usepackage{pxfonts,txfonts}

\newcommand{\bq}{\begin{equation}}
\newcommand{\eq}{\end{equation}}
\newcommand{\bqs}{\begin{equation*}}
\newcommand{\eqs}{\end{equation*}}
\newcommand{\ba}{\begin{array}}
\newcommand{\ea}{\end{array}}
\newcommand{\bas}{\begin{array*}}
\newcommand{\eas}{\end{array*}}
\newcommand{\bqa}{\begin{eqnarray}}
\newcommand{\eqa}{\end{eqnarray}}
\newcommand{\bqas}{\begin{eqnarray*}}
\newcommand{\eqas}{\end{eqnarray*}}

\DeclareMathOperator{\tr}{Tr}

\newcommand{\ran}{\rangle}
\newcommand{\lan}{\langle}

\newcommand{\D}{\mathcal{D}}

\newcommand{\qed}{\nobreak \ifvmode \relax \else
\ifdim\lastskip<1.5em \hskip-\lastskip
\hskip1.5em plus0em minus0.5em \fi \nobreak
\vrule height0.3em width0.5em depth0.25em\fi}

\begin{document}


\title{Comment on ``Witnessed entanglement and the geometric measure of quantum discord"}
\author{Swapan Rana}
\email{swapanqic@gmail.com}
\author{Preeti Parashar}
\email{parashar@isical.ac.in}
\affiliation{Physics and Applied Mathematics Unit, Indian Statistical Institute, 203 B T Road, Kolkata, India}
\date{\today}

\begin{abstract} In a recent article [\href{http://dx.doi.org/10.1103/PhysRevA.86.024302}{Phys. Rev. A {\bf 86}, 024302 (2012)}], the authors have  derived some hierarchy relations between geometric discord and entanglement (measured by negativity and its square). We point out that these results are incorrect and give analytic counterexamples. We also discuss briefly the reason for such violations. 

\end{abstract}

\maketitle


We start with the definition of GD and negativity from Ref. \cite{DebarbaetalPRA12}. For an $m\otimes n$ ($m\le n$) state $\rho$, they have defined 
\begin{equation}
\label{dp}D_{(p)}(\rho)=\min_{\xi\in\Omega}\|\rho-\xi\|^p_{(p)}
\end{equation} where $\Omega$ is the set of zero-discord (or classical-quantum) states given by $\xi=\sum p_i|i\ran\lan i|\otimes\rho_i$ and $\|X\|_{(p)}$ is the Schatten $p$ norm given by $\|X\|_{(p)}=\{\tr[X^\dagger X]^{p/2}\}^{1/p}$. The negativity was defined by 

\begin{subequations}\label{dneg}\begin{align}
N(\rho)&=\|\rho^{T_A}\|_{(1)}-1\label{dneg1}\\
&=\max\{0,-\min_{0\le W^{T_A}\le I}\tr(W\rho) \}\label{dneg2}
\end{align}
\end{subequations}
where $\rho^{T_A}$ is the partial transposition of $\rho$ with respect to $A$ and $W$ is any optimal entangled witness. Attempting to prove a conjecture made in \cite{GirolamiAdessoPRA11}, they have claimed  (Eq. 17 therein) that all bipartite states satisfy
\begin{equation}\label{con1}
D_{(2)}\ge \frac{N^2}{(m-1)^2}
\end{equation}

We first observe a typo that the two definitions of negativity in Eq. \eqref{dneg} are not equal, as the quantity in Eq. \eqref{dneg2} is the sum of absolute values of negative eigenvalues of $\rho^{T_A}$ whereas that in Eq. \eqref{dneg1} is just double of it.

As a first gap in their derivation (which could be taken as another typo, though), we note that they have not normalized $D_{(2)}$ to have maximum value unity, as has been done in the original paper \cite{GirolamiAdessoPRA11}. As a result, if we take Eq. \eqref{dneg1} as definition of negativity, Eq. \eqref{con1} is not necessarily satisfied even by the two-qubit maximally entangled state (any one of the Bell states has $D_{(2)}=1/2$ whereas $N=1$). The importance of normalization could be found in Refs. \cite{PianiPRA12, ChitambarPRA12}.   Now we will try to remove all these (possible) typos and show that the relation \eqref{con1}, whether normalized or not, is always violated by some states. 

Let us first consider the case when $D_{(2)}$ is normalized (taking $\D=\frac{m}{m-1}D_{(2)}$) and Eq. \eqref{dneg1} is taken as the definition of negativity. Then it becomes the original conjecture ($\D\ge N^2/(m-1)^2$) made in Ref. \cite{GirolamiAdessoPRA11}, which we have refuted recently in Ref. \cite{RanaParasharAr12}. Note that the normalizing factor $m/(m-1)>1$  and hence this case also includes the case when $D_{(2)}$ is not normalized and Eq. \eqref{dneg1} is taken as the definition of negativity.  Now we will give an analytic example to show that there are states violating even the weaker relation \cite{illegal1}
\begin{equation}\label{con2}
\frac{m}{m-1}D_{(2)}\ge \frac{N^2}{(m-1)^2}=\frac{\left[\sum\limits_{\lambda_i<0}\lambda_i(\rho^{T_A})\right]^2}{(m-1)^2} 
\end{equation}

Consider the $m\otimes m$ Werner state given by \bq\label{wernerstate} \rho_w=\frac{m-z}{m^3-m}\mathbf{I}+\frac{mz-1}{m^3-m}F,\quad z\in[-1,1]\eq
where $F=\sum|k\ran\lan l|\otimes|l\ran\lan k|$ and set $m=8,z=-1$. Now, if we consider the matrix form  of $\rho_w$ (in computational basis) as the state of a $2\otimes 32$ system, the left hand side of Eq. \eqref{con2} becomes $1/49$ while the right hand side becomes $25/784$. Though we have used the formula   for $D_{(2)}$ developed in Ref. \cite{RanaParasharPRA12} (which is exact for $2\otimes n$ states), a measurement in computational basis will yield the result. Any value of $z\in [-1,-34/43)$ will also work well. We note that for large enough $n$, the Refs. \cite{PianiPRA12, TufarellietalAr12} give enough intuition for violation of this relation. Nonetheless, the analytic counterexample makes it more explicit. 

The authors of Ref. \cite{DebarbaetalPRA12} also proposed to take $D_{(1)}$ as a proper measure of geometric discord and derived the hierarchy relation (Eq. 27 therein)
\begin{equation}\label{con3}
D_{(1)}\ge N
\end{equation}
However, as can be seen easily, this result is also not correct. It is well known that the trace distance satisfies  $\|\rho-\sigma\|_{(1)}\le 2$, for all 
$\rho$ and $\sigma$ \cite{etalNielsenPRA05}. Hence we must have $D_{(1)}\le 2$ whereas $N$ can take value up to $(m-1)/2$ (for example consider the Bell state in $m\otimes m$). Taking the identity matrix as the classical-quantum state (need not be optimal), we see that the relation is violated by $4\otimes 4$ Bell states. 

As has been pointed out in Ref. \cite{PianiPRA12}, the violation for $D_{(2)}$ stems from the fact that the Hilbert-Schmidt norm is not monotone--- $D_{(2)}$ could be increased or decreased by adding or removing a factorized local ancilla. We would like to mention that the trace norm being monotone, does not suffer from this problem. Thus the proposal of taking $D_{(1)}$ as a good measure, is interesting and might be worth investigating. However, as we pointed out here, establishing any interrelation should be done carefully. Another point of concern regarding the use of the trace norm is that its analytic calculation is very difficult and hence the main spirit of usual geometric discord will be lost.

\textit{Note added:} After submission of this comment to journal, we found an erratum posted on arXive \cite{erratum} in which the authors have tried to fix the errors, based on our criticism. Though the erratum is out of purview of the present comment, we would like to point out that some further modifications are needed. First of all, there is no state $\rho$ for which $n_{\_}=d-1$ holds; so the relation must be a strict inequality. Although the relation [$D_{(2)}>N^2/(d-1)$] will then be correct for NPT states (the PPT states trivially satisfy equality), an important point is that it can no longer be used to compare geometric discord and entanglement. We note that, both $N$ and $N^2$ are entanglement monotones, but $N/d$ (in general $N/f(d)$) is not an entanglement monotone, as it might be increased with removal (or addition) of local ancillary systems. This observation applies equally well to the interrelation between $D_{(1)}$ and $N/d$.


\begin{thebibliography}{20}
\bibitem{illegal1} This relation is not well-justified for comparison, because the left hand side varies from $0$ to $1$ whereas the right hand side from $0$ to $1/4$ (i.e., not normalized). Note that a violation of Eq. \eqref{con2} automatically implies a violation of Eq. \eqref{con1}.
\end{thebibliography}
\end{document}